# Response Regimes in Equivalent Mechanical Model of Strongly Nonlinear Liquid Sloshing

*M. Farid and O. V. Gendelman[*]*

*Faculty of Mechanical Engineering, Technion – Israel Institute of Technology*

*\*contacting author, ovgend@tx.technion.ac.il*

## Abstract

We consider equivalent mechanical model of liquid sloshing in partially-filled cylindrical vessel both for free vibrations case and for horizontal harmonic ground excitation. The model treats both the regime of linear sloshing, and strongly non-linear sloshing regime; the latter is related to hydraulic impacts applied to the vessel walls. These hydraulic impacts are commonly simulated with the help of high-power potential and dissipation functions. For the sake of analytic exploration, we substitute this traditional approach by treatment of an idealized vibro-impact system with velocity-dependent restitution coefficient. Parameters of the vibro-impact model are derived from the high-power potential and dissipation functions. The obtained reduced model is similar to recently explored system of linear primary oscillator with attached vibro-impact energy sink. Analysis is based on a multiple – scale approach; the ratio of modal mass of the first sloshing mode to the total mass of the liquid and the tank serves as a natural small parameter. In the case of external ground forcing, steady-state responses and chaotic strongly modulated responses are revealed. Besides, the system response to horizontal periodic excitation with additional Gaussian white noise, and corresponding dynamics on the slow invariant manifold are explored. All analytical predictions of the reduced vibro-impact model are validated against direct numerical simulations of "initial" equivalent model with high-power smooth potential and dissipation functions, and good agreement is observed.

# 1. Introduction

Cylindrical vessels filled with liquid are used in many fields of engineering, including nuclear, vehicle and aerospace industries, for storage of chemicals, gasoline, water, and other possibly hazardous liquids. Oscillations of the liquid in the vessel (liquid sloshing) may be dangerous for the vessel safety. So far, detailed analytical explorations are limited to small-amplitude sloshing in rectangular and cylindrical vessels. While being most interesting and potentially hazardous, high-amplitude liquid sloshing in cylindrical tanks still lacks complete analytic description. High-amplitude sloshing can cause hydraulic jumps. In this case major hydraulic impacts can act on the vessel structure walls [1]. Hydraulic jumps and wave collisions with vessel shells are the source of strong non-linearities in the system.

In this paper we adopt the equivalent mechanical model of the high-amplitude sloshing. This well-known model simulates the effects of hydraulic impacts with the help of high-order smooth potential and damping functions [2–4], following Pilipchuk and Ibrahim [5]. These functions are suitable for numeric simulations, but hardly applicable for analytic treatment. In order to pursue the analytic approach, we further simplify the model and substitute the high-order potential and damping functions by inelastic vibro-impact interactions. So, the resulting model includes both linear coupling and the vibro—impact constraints. In many previous studies the impact-induced dissipation was modeled by Newton method, which invokes traditional constant restitution coefficient [6,7]. However, this assumption is not valid for high impact velocities [8], and, of course, not for fluids. In the current work we demonstrate that the high-power potential and damping model allows extension of the Newton approach. The effective restitution coefficient can be expressed as a function of impact velocity, material properties and tank geometry. Under these assumptions, the model turns out to be treatable analytically, similarly to recently explored dynamics of linear oscillator with attached vibro—impact nonlinear energy sink [7,9,10]. To validate the results, direct numeric simulations with high-power potential and damping functions are used.

In traditional phenomenological models, the sloshing dynamics in a partially-filled liquid tank is modeled by a mass-spring-dashpot system or a pendulum, when each sloshing mode is modeled by a different modal mass. One can easily understand that the former dynamics is less complex, since it involves only one- dimensional dynamics and interactions. In the same time, it fails to represent vertical liquid motion (e.g. water jets [11]) and vertical excitation, that are better represented by the pendulum model (parametric excitation of liquid-filled vessel modeled by high-exponent potential pendulum by El-Sayad [12], Pilipchuk and Ibrahim[5]). Moreover, mass-spring-dashpot system is more common among engineering design regulations, as shown by Malhotra et al. [13]. It is worth mentioning that parameter values for both models mentioned above are presented by Dodge [14] and Abramson [15].

Current study deals only with seismic-induced horizontal excitations and hence horizontal internal forces. Then, one can adopt that the motion of the liquid in the partially-filled liquid tank with total mass M is approximately described by the mass-spring-dashpot system with mass $m$, stiffness $k$, linear viscosity of $c$ and displacement $y$ with respect to the vessel centerline. In this simplified model, three dynamical regimes can take place,

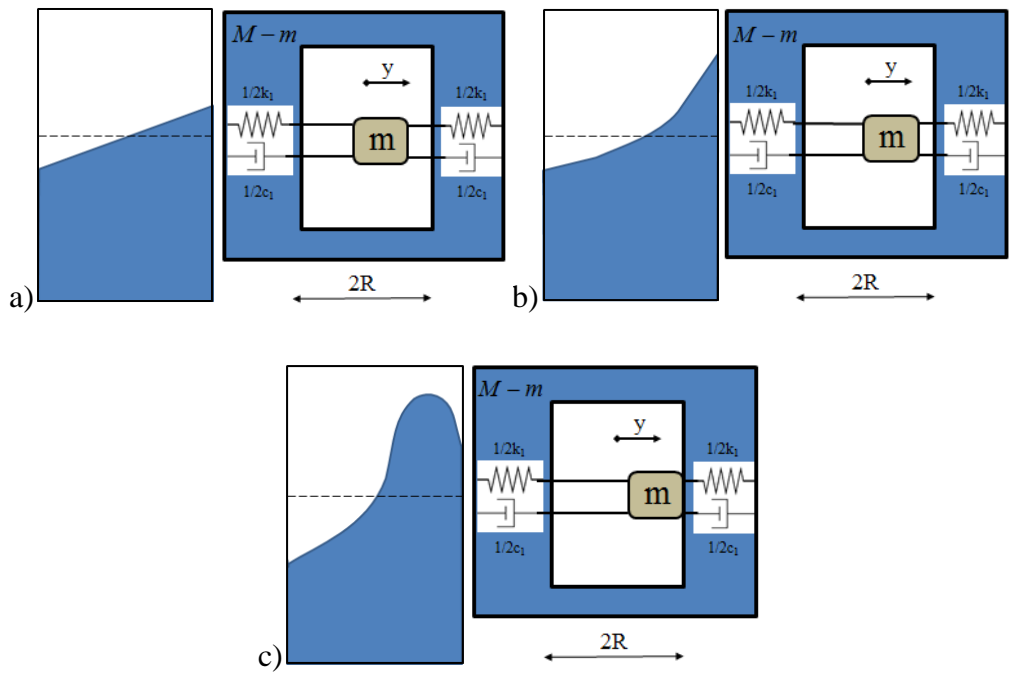

*Figure 1- Regimes of liquid free-surface motion and their equivalent mechanical models.*

(a) The liquid free surface performs small oscillations around its trivial stable equilibrium and remains planar. This regime can be successfully described by a linear mass-spring-dashpot system performing small oscillations.

(b) Relatively large oscillations, in which the liquid free surface does not remain planar. This motion is described by differential equation with weak nonlinearity, and can be treated by perturbation methods [5,16,17]. In this regime the equivalent mass-spring-dashpot system performs moderate oscillations, so that a cubic correction to the spring stiffness is reasonable, and the nonlinearity can be treated as weak.

(c) The free liquid surface is urged into a strongly nonlinear motion, related to liquid sloshing impacts with the tank walls. This regime can be described with the help of a mass-spring-dashpot system, which impacts the tank walls.

In the current study we use lumped mass *m* to model both linear liquid sloshing and nonlinear hydraulic impact regimes of the first asymmetric sloshing mode. Lumped mass-spring-damper model is also used to mimic the first bending mode of the vessel. Dynamical responses in the weakly nonlinear case are considered in previous work [18].

This paper is organized as follows: in section 2 we introduce the model, develop an analytical model for the restitution coefficient as a function of the impact velocity, and formulate the governing equations of motion. In section 3, asymptotic approximation for 1:1 internal resonance regime both for free vibrations and for the case of periodic forcing is developed. In section 4 numerical validation of the asymptotic results is presented and effect of additional stochastic excitation on the system response is explored.

## 2. Description of the model

**2.1. Introducing the model.**

Following the arguments presented in the Introduction, the model comprises damped linear oscillator (with possible external forcing), with internal vibro-impact particle with masses of M and m, respectively. Scheme of the system is presented in Figure 2. Absolute displacements of the primary mass (PM) and impacting particle (IP) are denoted as $u(t)$ and $v(t)$ respectively. The PM linear stiffness and viscosity are denoted as $\bar{k}$ and $\bar{c}$, respectively.

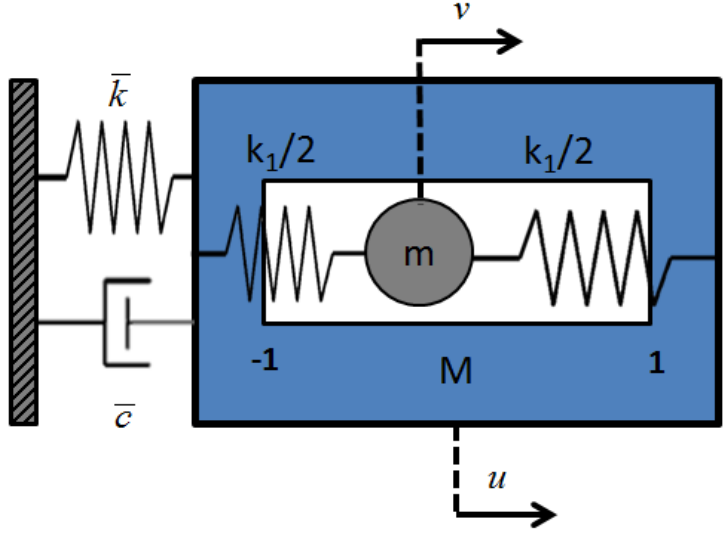

*Figure 2- Sketch of the system - linear oscillator as the primary system and internal vibro-impact particle with additional linear springt*

The IP is located inside a straight cavity in the PM, and in contrast to earlier explored vibro-impact NES [7,9,19], it is attached to it through a linear spring with stiffness $k_1$ that represents the liquid first sloshing mode mass $m$ and natural frequency $\omega = \sqrt{k_1/m}$. Without loss of generality, we consider the cavity length to be equal to 2. The linear spring is required to mimic the oscillatory sloshing motion due to gravity and recoiling from the vessel inner walls after impact. The external forcing at this stage is considered to be harmonic, with frequency $\Omega$ and amplitude $\varepsilon M \bar{A}$.

## 2.2. Equations of motion

The IP relative displacement with respect to the PM is defined as:

$$w(t) = u(t) - v(t) \qquad (1)$$

Impact takes place for $|w(t_j)| = 1$, where $t_j$ is the impact event instant. The impact effect is described with the help of a restitution coefficient $\kappa$ which in general depends on the impact velocity:

$$w(t_j^+) = -\kappa\left(w(t_j^-)\right) w(t_j^-) \qquad (2)$$

Here $t_j^-$ and $t_j^+$ are the time instants before and after the impact, respectively. Newtonian approach is a particular case of relation (2) for which $\kappa$ is constant. This assumption is valid only for small velocities in soft materials and for intermediate velocities for hard materials. In current study we use more general relation between the restitution coefficient $\kappa$ and impact velocity $w(t_j^-)$ (see below, Section 2.3).

Momentum conservation in the course of impact yields the following relation:

$$M\dot{u}(t_j^-) + m\dot{v}(t_j^-) = M\dot{u}(t_j^+) + m\dot{v}(t_j^+) \tag{3}$$

From equations (1)-(3), the total amount of momentum transmitted to the PM in each impact is expressed as follows:

$$\Delta P = M\left(\dot{u}(t_j^+) - \dot{u}(t_j^-)\right) = \frac{-Mm}{M+m}\left(\kappa\left(w(t_j^-)\right) + 1\right)\dot{w}(t_j^-) \tag{4}$$

Hence, the normalized equations of motion obtain the following form::

$$\begin{aligned}\ddot{u} + \varepsilon\zeta\dot{u} + \omega^2 u + \frac{\varepsilon(\kappa+1)}{1+\varepsilon}\sum_j \dot{w}(t_j^-)\delta(t-t_j) &= \varepsilon\bar{A}\cos(\Omega t) \\ \ddot{v} + \omega_1^2 v - \frac{\kappa+1}{1+\varepsilon}\sum_j \dot{w}(t_j^-)\delta(t-t_j) &= 0\end{aligned} \tag{5}$$

Here $\varepsilon = m/M$ is the IP and PM mass ratio, $\omega_1 = \sqrt{k_1/m}$ is the IP natural frequency and $\omega = \sqrt{k/M}$, $\varepsilon\zeta = \bar{c}/M$ are the PM natural frequency, and damping coefficient, respectively. $\delta(t)$ is a Dirac delta function. Calculation of realistic values of these parameters will be presented in details elsewhere [20].

We introduce coordinate which is proportional to the displacement of the center-of-mass, and define the following non-dimensional parameters and time rescaling:

$$\begin{aligned}X(t) &= u(t) + \varepsilon v(t) \\ \beta &= \frac{\omega_1}{\omega}; \quad \gamma = \frac{\zeta}{\omega\sqrt{(1+\varepsilon)(1+\varepsilon\beta^2)}}; \quad \tau = \omega\sqrt{\frac{1+\varepsilon\beta^2}{1+\varepsilon}}t\end{aligned} \tag{6}$$

Here $\beta$ and $\gamma$ are the ratio of natural frequencies and the non-dimensional damping coefficient, respectively. Further analysis will focus on the most interesting case of primary 1:1 resonance. Therefore a non-dimensional detuning parameter $\sigma$ is introduced. We implement time normalization and coordinates transformation of equations (5) according to equations (1) and (6) to obtain the non-dimensional equations of motion:

$$\ddot{X} + (1-\varepsilon\sigma)X + \frac{\varepsilon(1-\beta^2)(1-\varepsilon\sigma)}{1+\varepsilon\beta^2}w + \varepsilon\gamma\dot{X} + \varepsilon^2\gamma\dot{w} = \varepsilon\bar{A}\cos\tau$$

$$\ddot{w} + \frac{(1-\beta^2)(1-\varepsilon\sigma)}{1+\varepsilon\beta^2}X + \frac{(\varepsilon+\beta^2)(1-\varepsilon\sigma)}{1+\varepsilon\beta^2}w + \varepsilon\gamma\dot{X} + \varepsilon^2\gamma\dot{w} + \sum_j\left(1+\kappa\left(\dot{w}(\tau_j^-)\right)\right)\dot{w}(\tau_j^-)\delta(\tau-\tau_j^-) = \varepsilon\bar{A}\cos\tau$$

(7)

The dot stands for derivation with respect to normalized time $\tau$. We assume that $\beta$ is of order unity, and expand the coefficients presented in equations (7) to Taylor series up to order of $\varepsilon$ to yield the system equations of motion in their final form:

$$\ddot{X} + (1-\varepsilon\sigma)X + \varepsilon(1-\beta^2)w + \varepsilon\gamma\dot{X} = \varepsilon\bar{A}\cos\tau$$

$$\ddot{w} + \left((1-\beta^2)-\varepsilon(1-\beta^2)(\sigma+\beta^2)\right)X + \left(\beta^2 + \varepsilon\left((1-\beta^2)-\sigma\beta^2\right)\right)w +$$

$$+ \varepsilon\gamma\dot{X} + \sum_j\left(1+\kappa\left(\dot{w}(\tau_j^-)\right)\right)\dot{w}(\tau_j^-)\delta(\tau-\tau_j^-) = \varepsilon\bar{A}\cos\tau$$

(8)

If, instead of the vibro-impact interaction, one uses the high-power smooth potential and dissipation functions to model the inelastic hydraulic impacts, system (8) is modified as follows:

$$\ddot{X} + (1-\varepsilon\sigma)X + \frac{\varepsilon(1-\beta^2)(1-\varepsilon\sigma)}{1+\varepsilon\beta^2}w + \varepsilon\gamma\dot{X} + \varepsilon^2\gamma\dot{w} = \varepsilon\bar{A}\cos\tau$$

$$\ddot{w} + \frac{(1-\beta^2)(1-\varepsilon\sigma)}{1+\varepsilon\beta^2}X + \frac{(\varepsilon+\beta^2)(1-\varepsilon\sigma)}{1+\varepsilon\beta^2}w + \varepsilon\gamma\dot{X} + \varepsilon^2\gamma\dot{w} + \lambda\dot{w}w^{2p} + \chi w^{2n+1} = \varepsilon\bar{A}\cos\tau$$

(9)

The relationship between parameters of Systems (8) and (9) will be established in the next Section.

### 2.3. Velocity-dependent restitution coefficient.

In order to establish the relationship between the non-elastic VI system and high-power smooth potential and dissipation functions, we evaluate the velocity-dependent restitution coefficient. For this sake, following [4,5,12], the equation of motion of a free particle with mass $\bar{m}$ between two rigid constraints located in $x = \pm R$ is considered in the following forms:

$$m\ddot{x} + \bar{\alpha}\left(\frac{x}{R}\right)^{2n+1} + \bar{\beta}\dot{x}\left(\frac{x}{R}\right)^{2p} = 0 \qquad (10)$$

Here $\bar{\alpha}$, $\bar{\beta}$ are phenomenological constants which are determined experimentally, and $n$ and $p$ are integers, significantly larger than unity and not necessarily equal. Corresponding potential energy, total energy and Rayleigh dissipation function $U(x)$, $E(x,\dot{x})$ and $D(x,\dot{x})$ may be written as follows:

$$U(x) = \frac{\bar{\alpha}R}{2(n+1)}\left(\frac{x}{R}\right)^{2(n+1)}, \quad E(x,\dot{x}) = \frac{\bar{m}}{2}\dot{x}^2 + U(x), \quad D(x,\dot{x}) = \bar{\beta}\dot{x}^2\left(\frac{x}{R}\right)^{2p} \qquad (11)$$

From equation (11) the energy balance condition is obtained:

$$\frac{d}{dt}E(x(t),\dot{x}(t)) = -D(x(t),\dot{x}(t)) \qquad (12)$$

As mentioned above, we are interested in estimating the dissipation related to collision in terms of velocity-dependent restitution coefficient, satisfying the following condition:

$$\kappa(v_0) = -v_+/v_0 \qquad (13)$$

Here $v_0$ and $v_+$ are the particle velocity in the equilibrium point before and after the collision, respectively. Obviously, $\kappa < 1$ due to energy dissipation.

Equation (12) is integrated between time zero and $t_c$, where $t_c$ is the time interval that the particle spends between subsequent passages through point $x = 0$:

$$\frac{\bar{m}}{2}(1-\kappa^2)v_0^2 = \int_{t=0}^{t_c} D(x,\dot{x})\,dt \qquad (14)$$

The integral in equation (14) can be separated into two integrals, corresponding to pre and post-impact motions:

$$\int_{t=0}^{t_c} D(x,\dot{x})\,dt = \bar{\beta}\left(\int_{x=0}^{x_m}\dot{x}_-\left(\frac{x}{R}\right)^{2p}dx - \int_{x=0}^{x_m}\dot{x}_-\left(\frac{x}{R}\right)^{2p}dx\right) \qquad (15)$$

Here $\dot{x}_-(x)$ and $\dot{x}_+(x)$ are pre- and post-impact velocities, respectively. In the case of relatively small dissipation, all values in the right-hand side of (8) can be evaluated from a conservative approximation:

$$\dot{x}_+(x) \approx \dot{x}_-(x) = v_0 \sqrt{1 - \left(\frac{x}{x_{\max}(v_0)}\right)^{2(n+1)}} \qquad (16)$$

Here $x_{\max}(v_0)$ is the maximum displacement of the particle in the external high-power potential with given initial velocity of $v_0$. For general $n$, the following relationship holds:

$$x_{\max}(v_0) = R \left(\frac{\bar{m}(n+1)}{\bar{\alpha} R}\right)^{\frac{1}{2(n+1)}} v_0^{\frac{1}{n+1}} \qquad (17)$$

It is noteworthy that for perfect square potential well, corresponding to $n \to \infty$, we get $x_{\max} = R$. Substitution of (16) into (15) yields:

$$\int_{t=0}^{T/2} D(x, \dot{x}) dt \approx 2\bar{\beta} R \left(\frac{\bar{m}(n+1)}{\bar{\alpha} R}\right)^{\frac{2p+1}{2(n+1)}} I(n,p) v_0^{\frac{2p+n+2}{n+1}}; \quad I(n,p) = \int_0^1 \zeta^{2p} \sqrt{1 - \zeta^{2(n+1)}} d\zeta \qquad (18)$$

Defining new parameter $\bar{p} = 2p - n$, one obtains:

$$I(n, \bar{p}) = \frac{\sqrt{\pi} \Gamma\left(\frac{\bar{p}}{2(n+1)} + \frac{1}{2}\right)}{4(n+1) \Gamma\left(\frac{\bar{p}}{2(n+1)} + 2\right)} \qquad (19)$$

Dependence of integral $I(n, \bar{p})$ on parameters n and $\bar{p}$ is illustrated in Figure 3. Expression (19) is valid for $\forall n \in \mathbb{N}: \bar{p} > -(n+1)$. This fact does not lead to loss of generality, since in terms of parameter $p$ it means $p > -1/2$, i.e. expression (19) is valid for every feasible value of $p$.

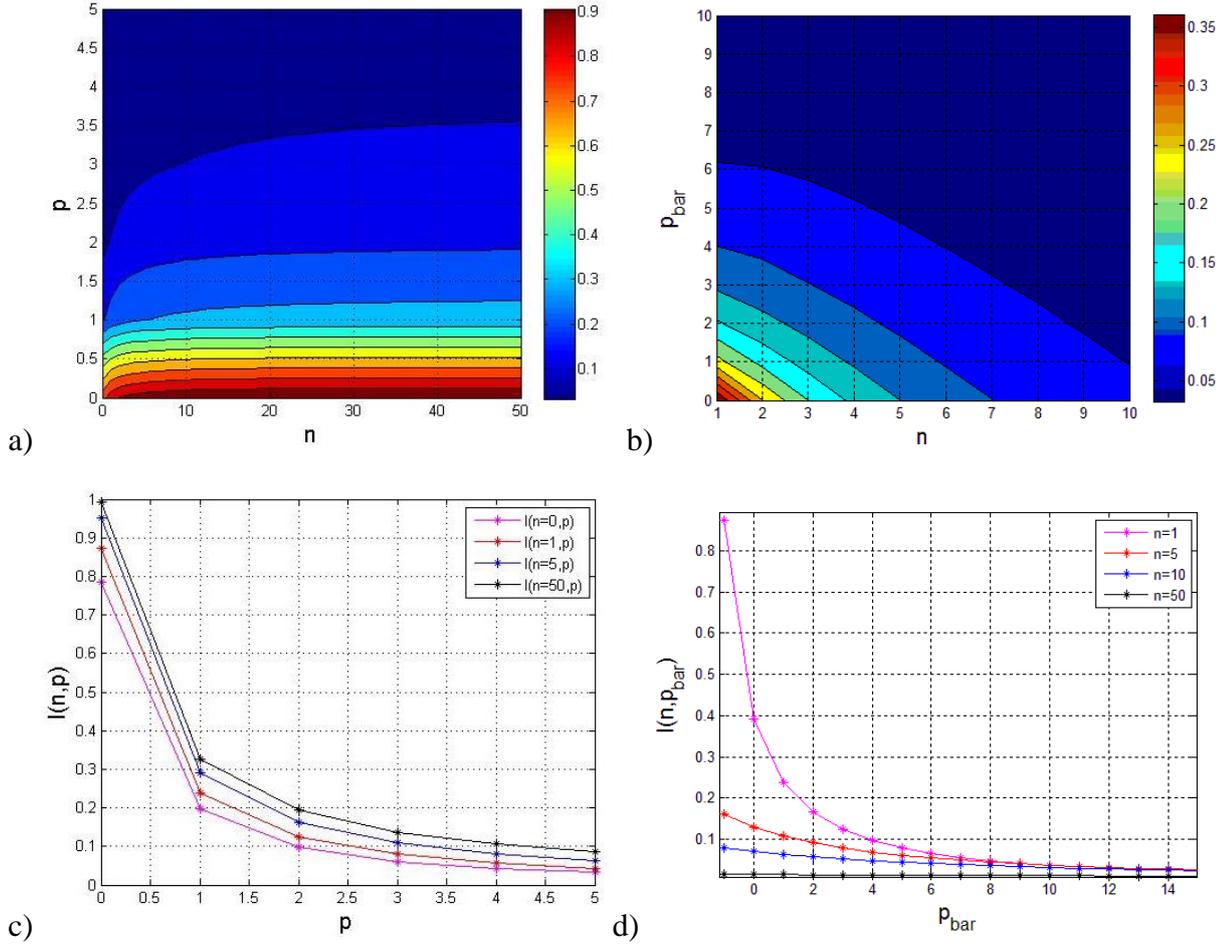

*Figure 3- a),b) Plot of integrals $I(n,p)$ and $I(n,\bar{p})$ for parameter range $n \in [0,50]$ and $p \in [0,5]$ ;c),d) Plot of integrals $I(n,p)$ and $I(n,\bar{p})$ vs. parameters n and for parameter range c) Plots of integral $I(n,p)$ and $I(n,\bar{p})$ vs. parameter $p \in [0,20]$ ; for n=0 (pink), n=1(red), n=5(blue)and n=50(black).d)zoom-in for $p \in [0,5]$ .*

For the Newtonian case:

$$I(n,\bar{p}=0) = \frac{\sqrt{\pi}\Gamma\left(\frac{1}{2}\right)}{4(n+1)\Gamma(2)} = \frac{\pi}{4(n+1)} \tag{20}$$

As expected, for $\bar{p} = 0$ and large values of parameter *n* integral $I(n,\bar{p})$ tends to zero. From Figure 3 we can learn that for any selection of parameter $\bar{p}$, the value of integral $I(n,\bar{p})$ decreases with increasing value of parameter *n*.

Substituting (18) into(14), one obtains the relationship between restitution coefficient $\kappa$, the pre-impact velocity and the parameters of high-power force and damping :

$$\kappa(v_0, n, \bar{p}) = \sqrt{1 - 4\lambda \left(\frac{n+1}{\chi}\right)^{\frac{\bar{p}}{2(n+1)} + \frac{1}{2}} I(n, \bar{p}) v_0^{\frac{\bar{p}}{n+1}}} \qquad (21)$$

$$\lambda = \frac{\bar{\beta} R}{m}, \quad \chi = \frac{\bar{\alpha} R}{m}$$

Parameters $\lambda, \chi, n$ and $\bar{p}$ are determined experimentally. One can learn from equation (21), that when integral $I(n, p)$ tends to zero, the dependence of the restitution coefficient $\kappa$ on pre-impact velocity $v_0$ becomes weaker and $\kappa$ tends to unity. One can see that selection of parameter $n$ for theoretical analysis cannot be arbitrary and should satisfy physical requirements of real-positive and smaller than unity restitution coefficient.

## 3. Multiple-scale analysis.

The modal masses of the asymmetric sloshing modes in cylindrical vessels were well documented by Abramson [15]. Figure 4 demonstrates the dependence of the 'static' fluid portion ratio $m_0/m_F$, and modal sloshing mass ratios of the first three sloshing modes versus the depth-radius ratio $h/r$. One can see that all modal masses decrease rapidly with fluid depth. Besides, it is obvious that the effective mass of the first sloshing mode is most significant.

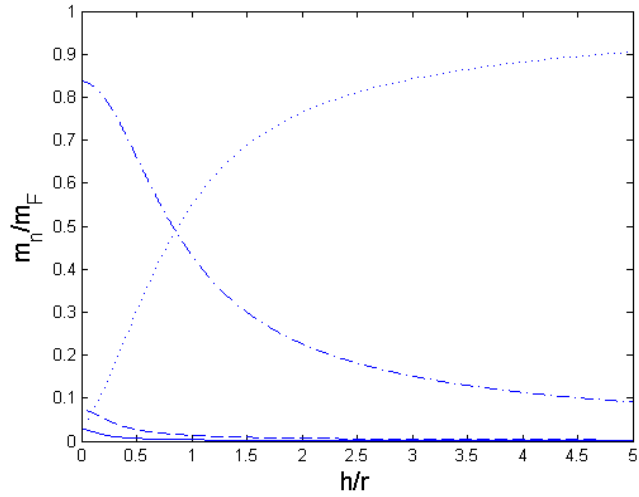

*Figure 4-Ratios of the first three asymmetric sloshing modal masses $m_1$, $m_2$ and $m_3$ and fixed mass $m_0$ to the total fluid mass $m_F$ for cylindrical vessel; dotted-lime: $m_0/m_F$, dashed-dotted-line: $m_1/m_F$ ,dashed-line: $m_2/m_F$ solid-line: $m_3/m_F$.*

Consequently, we take into account only the first sloshing mode; it is possible to adopt $m \approx m_1$. The "inert" mass $M$ includes both the non-sloshing liquid portion $m_0$ and the mass of the vessel shell $M = m_0 + m_{\tan k}$.

In view of the above findings, in further analysis the mass ratio is considered to be rather small, $\varepsilon = m/M \ll 1$. All other parameters are considered to be of order unity. We introduce multiple scales approach, using the following expansions:

$$T_k = \varepsilon^k \tau; \quad k = 0,1,...$$
$$D_k = \frac{d}{dT_k}; \quad \frac{d}{d\tau} = D_0 + \varepsilon D_1 + ... \quad (22)$$
$$X = X_0(T_0, T_1) + \varepsilon X_1(T_0, T_1)$$
$$w = w_0(T_0, T_1) + \varepsilon w_1(T_0, T_1)$$

In the current problem, only the fast time scale $T_0$ and the slow time scale $T_1$ are necessary. Substitution of expansions (22) into equations (8) yields:

$$\left(D_0^2 + 2\varepsilon D_0 D_1\right)(X_0 + \varepsilon X_1) + (1 - \varepsilon\sigma)(X_0 + \varepsilon X_1) + \varepsilon(1-\beta^2)(w_0 + \varepsilon w_1) +$$
$$+ \varepsilon\gamma(D_0 + \varepsilon D_1)(X_0 + \varepsilon X_1) = \varepsilon \bar{A}\cos(T_0) \quad (23)$$
$$\left(D_0^2 + 2\varepsilon D_0 D_1\right)(w_0 + \varepsilon w_1) + \left((1-\beta^2) - \varepsilon(1-\beta^2)(\sigma + \beta^2)\right)(X_0 + \varepsilon X_1)$$
$$+ \left(\beta^2 + \varepsilon((1-\beta^2) - \sigma\beta^2)\right)(w_0 + \varepsilon w_1) + \varepsilon\gamma(D_0 + \varepsilon D_1)(X_0 + \varepsilon X_1)$$
$$+ \sum_j \left(1 + \kappa\left((D_0 + \varepsilon D_1)(w_0 + \varepsilon w_1)\right)\right)(D_0 + \varepsilon D_1)(w_0 + \varepsilon w_1)\delta(T_0 - T_{0j}^-) = \varepsilon\bar{A}\cos(T_0)$$

We collect term of $O(1)$ in equation (23) and obtain the following fast time-scale equations:

$$D_0^2 X_0 + X_0 = 0$$
$$D_0^2 w_0 + \beta^2 w_0 + (1-\beta^2)X_0 + \sum_j \left(1 + \kappa\left(D_0 w_0(T_{0,j}^-)\right)\right)D_0 w_0(T_{0,j}^-)\delta(T_0 - T_{0,j}) = 0 \quad (24)$$

The first equation in system (24) corresponds to the free linear oscillator with natural frequency of unity:

$$X_0 = A(T_1)\sin(T_0 + \psi(T_1)) \quad (25)$$

Here $A(T_1)$ and $\psi(T_1)$ are slowly varying amplitude and phase of $X$, respectively. The solution of the second equation in system (24) is obtained by substituting (25) into the second equation of (24). It consists of particular solution, homogeneous solution, and a third term $f(T_0, T_1)$ corresponding to the impact term:

$$w_0 = B(T_1)\sin(\beta T_0 + \psi(T_1)) + A(T_1)\sin(T_0 + \phi(T_1)) + f(T_0, T_1) \quad (26)$$

Here the first term in (26) is the homogeneous solution of equation (24), and the other terms are the heterogeneous solutions corresponding to two heterogeneous terms in equation (24).

As mentioned above, the restitution coefficient depends on the velocity at the impact instant: $\kappa\left(\left|D_0 w_0\left(T_{0,j}^-\right)\right|\right)$.

$$\left|D_0 w_0\left(T_{0,j}^-\right)\right| = \left|\beta B(T_1)\cos(\beta T_0 + \phi) + C(T_1)\cos(T_0 + \psi) + \alpha\right|$$

$$\begin{cases} w=1: & \left|D_0 w_0\left(T_0 = (\eta + 2s\pi)_-\right)\right| = \left|\beta B(T_1)\cos(\beta\eta + \phi) + C(T_1)\cos(\eta + \psi) + \alpha\right| \\ w=-1: & \left|D_0 w_0\left(T_0 = (\eta + (2s+1)\pi)_-\right)\right| = \left|-\beta B(T_1)\cos(\beta\eta + \phi) - C(T_1)\cos(\eta + \psi) - \alpha\right|, \quad s=0,1,2,\ldots \end{cases} \quad (27)$$

$$\to \left|D_0 w_0\left(T_0 = (\eta + 2s\pi)_-\right)\right| = \left|D_0 w_0\left(T_0 = (\eta + (2s+1)\pi)_-\right)\right|$$
$$\to \kappa(T_0 = \eta + 2s\pi) = \kappa(T_0 = \eta + (2s+1)\pi) \equiv \kappa(T_1)$$

As one can see from equation (27), in the lowest-order approximation, the absolute value of impact velocity at both boundaries is equal. Hence, the restitution coefficient should be considered constant with respect to the fast time scale.

Substituting (25)-(27) to (24)(b), one obtains:

$$D_0^2 f + \beta^2 f - \beta^2(1-C)\sin(T_0+\psi) + \sum_j (1+\kappa)\left(D_0 f(T_0^-) + \beta B\cos(\beta T_0 + \phi)C\cos(T_0+\psi)\right)\delta(T_0 - T_{0,j}) = 0 \quad (28)$$

The most common and interesting response corresponds to 1:1 resonance between IP impacts and PM oscillations. Hence, we look for solution of (28), which corresponds to the steady state periodic motion of a particle between two rigid berries subjected to external excitation. The oscillation amplitude will be constant with respect to the fast time scale $T_0$. The solution in general form may be presented as follows [21–23]:

$$f(T_0, T_1) = \frac{2\alpha(T_1)}{\pi}\arcsin(\cos(T_0 - \eta)) \quad (29)$$

Here $\alpha$ is the slowly-varying amplitude of function $f(T_0, T_1)$ and $\eta$ is its phase shift with respect to the excitation force, as one can see in Figure 5 ($\alpha$ is constant with respect to fast time scale $T_0$).

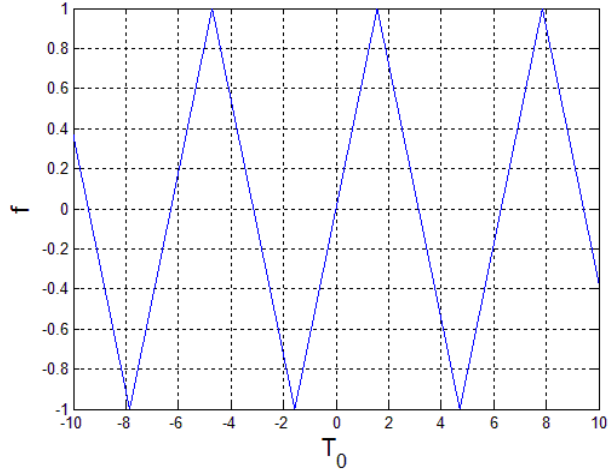

*Figure 5- Plot of $f(T_0)$, for $\alpha = 1$ and $\eta = \pi/2$*

Since we look for solution corresponding to the 1:1 resonance for which impact occurs twice in every oscillation period (once in every cavity wall), the impacts take place for $T_0 = \eta + (1-\bar{n})\pi$ where $\bar{n} = 1, 2, ...$ represents the impact number.

Substituting equation (29) into equation(28) and integrating over very small time interval around the instance of impact, one obtains::

$$\frac{4\alpha(T_1)}{\pi} - \frac{2\alpha(T_1)(1-\kappa(T_1))}{\pi} = (1+\kappa(T_1))(\beta B(T_1)\cos(\beta\eta + \phi(T_1)) + A(T_1)\cos(\eta + \psi(T_1))) \quad (30)$$

From (30), we derive the following equation:

$$\beta B(T_1)\cos(\beta\eta + \phi(T_1)) + A(T_1)\cos(\eta + \psi(T_1)) = \rho(T_1)\alpha(T_1); \quad \rho(T_1) = \frac{2(1-\kappa(T_1))}{\pi(1+\kappa(T_1))} \quad (31)$$

To satisfy the impact condition at $T_0 = \eta$, one should equate (26) to unity at this instance. This leads to the following equation:

$$B(T_1)\sin(\beta\eta + \phi(T_1)) + A(T_1)\sin(\eta + \psi(T_1)) = 1 - \alpha(T_1) \quad (32)$$

In the same way, we apply (26) for the impact at $w_0 = -1$ and use trigonometric identities to obtain:

$$\beta B(T_1)\cos(\beta\pi)\cos(\beta\eta+\phi(T_1)) - \beta B(T_1)\sin(\beta\pi)\sin(\beta\eta+\phi(T_1)) - A(T_1)\cos(\eta+\psi(T_1)) = -\rho(T_1)\alpha(T_1); \quad (33)$$
$$B(T_1)\cos(\beta\pi)\sin(\beta\eta+\phi(T_1)) + B(T_1)\sin(\beta\pi)\cos(\beta\eta+\phi(T_1)) - A(T_1)\sin(\eta+\psi(T_1)) = -(1-\alpha(T_1));$$

Solution of system(31)-(33) for $\sin(\beta\eta+\phi(T_1)), \cos(\beta\eta+\phi(T_1))$ $\sin(\eta+\psi(T_1))$ and $\cos(\eta+\psi(T_1))$ yields the following set of equations:

$$\begin{pmatrix} \sin(\eta+\psi(T_1)) \\ \cos(\eta+\psi(T_1)) \\ \sin(\beta\eta+\phi(T_1)) \\ \cos(\beta\eta+\phi(T_1)) \end{pmatrix} = \begin{pmatrix} \dfrac{1-\alpha}{A} + Bf_1(\alpha,\rho,\beta,A) \\ \dfrac{\rho\alpha}{A} + Bf_2(\alpha,\rho,\beta,A) \\ 0 \\ 0 \end{pmatrix} \quad (34)$$

Since both $\sin(\beta\eta+\phi(T_1))$ and $\cos(\beta\eta+\phi(T_1))$ cannot be equal to zero, we conclude that $B(T_1)=0$. It is noteworthy that this result is approximately valid also for resonant 1:1 oscillations regime when $\beta>0$, since DOF $w(t)$ becomes a linear oscillator due to irrelevance of the impact high-order terms. Following equations (26)and(29) the fast time scale solution of $w$ is as follows:

$$w_0(T_0,T_1) = A(T_1)\sin(T_0+\phi(T_1)) + \frac{2\alpha(T_1)}{\pi}\arcsin(\cos(T_0-\eta)) \quad (35)$$

Hence, the slow invariant manifold (SIM [24,25]) of the system is obtained from equations (31) and (32) as $B(T_1)$ is set to zero, as follows:

$$\alpha(A) = \frac{1\pm\sqrt{(1+\rho^2)(A^2-A_{min}^2)}}{1+\rho^2}; \quad A_{min}(T_1) = \frac{\rho(T_1)}{\sqrt{1+\rho(T_1)^2}} \quad (36)$$

Here $A_{min}$ is the minimum value possible for $X_0$ amplitude in conditions of the 1:1 resonance. The SIMconsists of a sole stable branch and unstable branch. The resulting dynamics can exhibit alternating fast captures into 1:1 resonance regime and rapid amplitude decay for varying amplitude and duration, separated by intervals of non-resonant dynamics. Hence, this regime exhibits chaotic characteristics, and is referred to aschaotic strongly modulated response (CSMR, [7]). Note that parameter $\rho$ depends on $\kappa(T_1)$. The first expression for $\alpha(A)$, denoted by $\alpha_{S^+}(A)$ and corresponding to the positive sign, represents the SIM stable branch (upper branch,

Figure 6) defined as $S^+ = \{A \in [A_{min}, \infty), \phi \in [0, 2\pi)\}$, while the other one represents the unstable branch (lower branch, Figure 6). Escape from the resonance, i.e. a "jump" from the SIM stable branch $S^+$ takes place when the response amplitude reaches $A = A_{min}$ and on the SIM bifircation point $\alpha_{min} = \alpha(A = A_{min}) = 1/(1+\rho^2)$. In other words, $A_{min}$ is the minimum $X_0$ amplitude for which 1:1 resonance is possible.

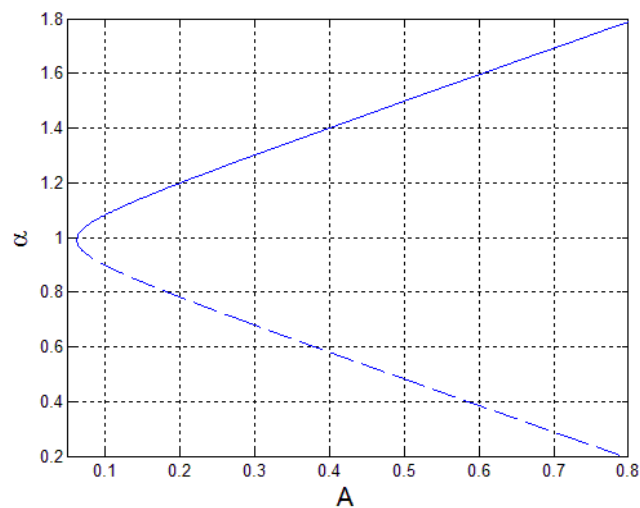

*Figure 6-The system SIM, for $\kappa = 0.729$, $\beta = 0.2$ and $A_{min} = 0.06$, $\alpha_{min} = 0.99$. Solid line: stable branch, dashed line: unstable branch.*

The restitution coefficient $\kappa$ is a function of the impact velocity; the latter is derived from expression (35) as follows:

$$|D_0 w_0 (T_0 = \eta_-)| = |A\cos(T_0 + \psi) + sgn(\sin(T_0 - \eta))|_{T_0 = \eta_-} \tag{37}$$

Taking into account (31), we get:

$$|D_0 w_0 (T_0 = \eta)| = \alpha \left(\frac{2}{\pi} + \rho\right) \tag{38}$$

Substitution of equations (38) and (31)(b) into (23) yields the following relationship:

$$\kappa(A) = \sqrt{1 - 4\lambda \left(\frac{n+1}{\chi}\right)^{\frac{\bar{p}}{2(n+1)} + \frac{1}{2}} I(n, \bar{p}) \left(\frac{4\alpha_{S^+}(A)}{\pi(1+\kappa(A))}\right)^{\frac{\bar{p}}{(n+1)}}} \tag{39}$$

From equation (39), the Newtonian case, corresponding to $\bar{p} = 0$, yields the following:

$$\kappa_{Newtonian}(A) = \sqrt{1 - \frac{\pi\lambda}{\sqrt{\chi(n+1)}}} \tag{40}$$

As expected, for the Newtonian case the restitution coefficient depends neither on the pre-impact velocity nor on the solution amplitude during impact regime. We substitute (36) to (39) to obtain the following relation between restitution coefficient $\kappa$ and amplitude $A$:

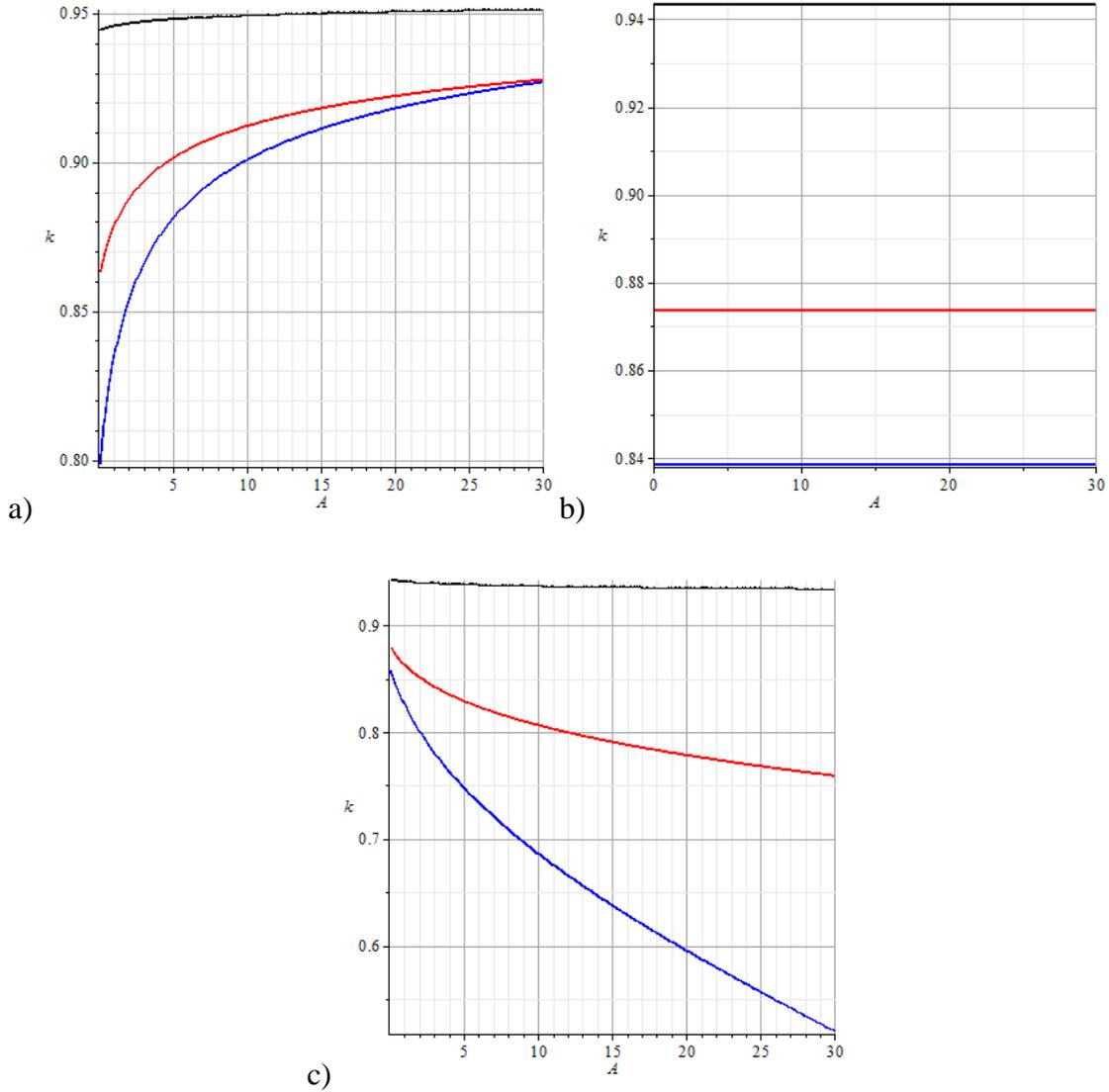

Figure 7- Restitution coefficient $\kappa$ vs. $X_0$ solution amplitude A for parameters:

$\lambda = 0.25, \chi = 1, n = 6(blue), n = 10(red), n = 50(black)$, for any value of $\beta$;

a) for $\bar{p} = -2$; b) $\bar{p} = 0$; c) $\bar{p} = 2$

As one can see, results corresponding to $\bar{p}=0$ shown in Figure 7 are in good agreement with the conclusions of [4] - $\bar{p}=0$ corresponds to the Newtonian impact, for which restitution coefficient $\kappa$ does not depend on the impact velocity. The corrected SIM diagram represented by equation (36), is shown in Figure 8.

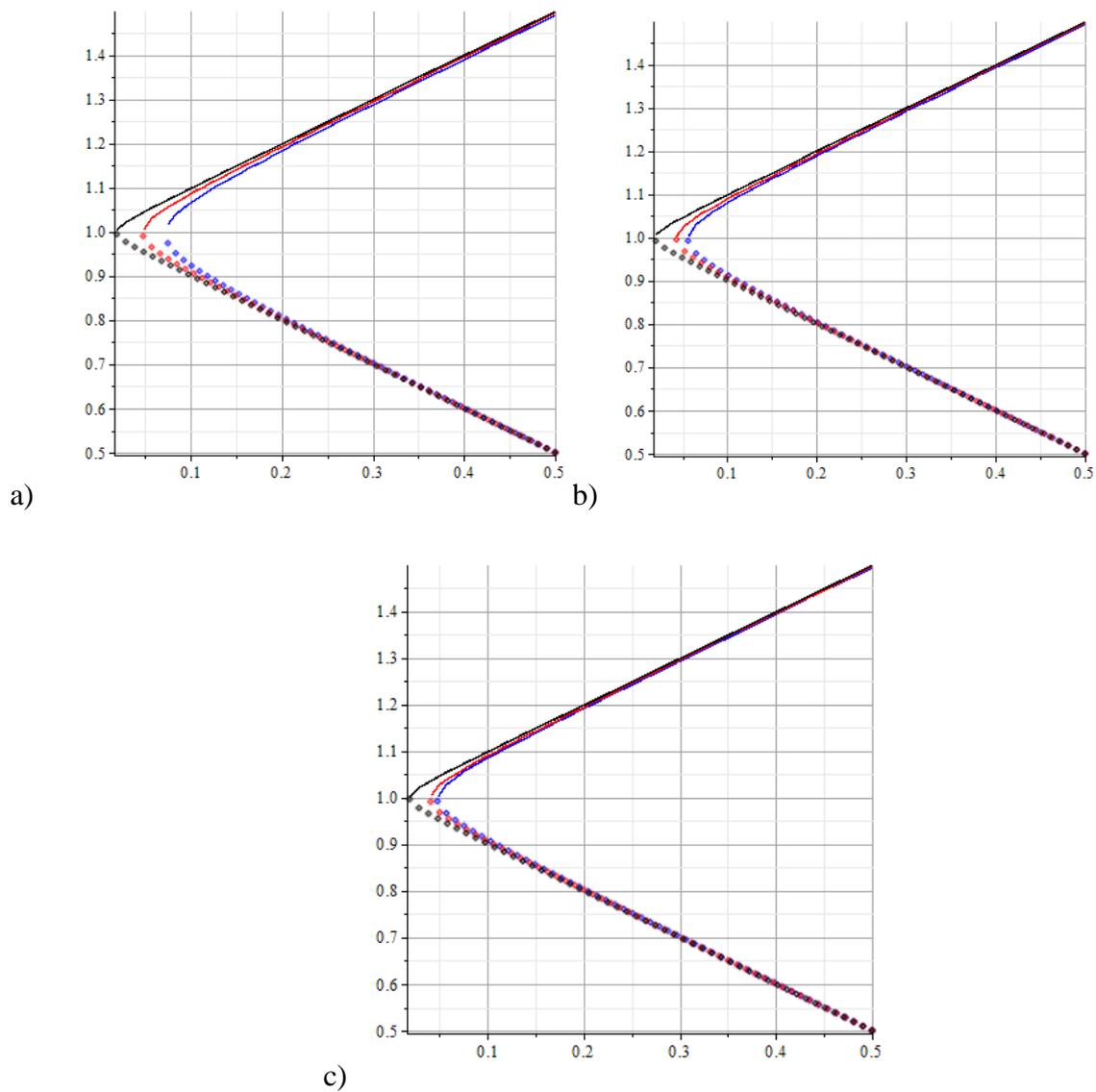

*Figure 8- The system SIM for various values of parameter $\bar{p}$,*
$\lambda = 0.25, \chi = 1, n = 6(blue), n = 10(red), n = 50(black)$, *for any value of* $\beta$;
*a)* r $\bar{p} = -2$; *b)* $\bar{p} = 0$; *c)* $\bar{p} = 2$

The analysis above is relevant for both free and forced vibration, since the excitation force with amplitude of order $\varepsilon$ does not affect the lowest-order.

## 3.1. Free vibrations

The slow evolution of amplitude on the SIM stable branch $S^+$ is described by equation corresponding to elimination of the secular term in slow time scale equations obtained from (8). Thus, we collect terms of order $\varepsilon$ to obtain the following slow-flow equation for amplitude $X$:

$$D_0^2 X_1 + X_1 = -2D_0 D_1 X_0 - \gamma(D_0 X_0) - (1-\beta^2) w_0 \qquad (41)$$

The function $f(T_0, T_1)$ is expanded into Fourier series as follows:

$$f(T_0, T_1) = \frac{2\alpha}{\pi} \arcsin(\cos(T_0 - \eta)) = \frac{8\alpha}{\pi^2} \sum_{\tilde{n}=1}^{\infty} \frac{(-1)^{\tilde{n}+1}}{(2\tilde{n}-1)^2} \cos((2\tilde{n}-1)(T_0 - \eta)) \qquad (42)$$

Then we combine expansion (42), the fast time scale solutions (25) and (35), and expression for the stable branch $\alpha_{S^+}$ in equation (36) with (41). Elimination of the secular terms leads to the following equations:

$$\begin{aligned} D_1 A &= -\frac{\gamma A}{2} - \frac{4\rho \alpha_{S^+}^2 (1-\beta^2) \rho}{\pi^2 A} \\ D_1 \psi &= \frac{(1-\beta^2)}{2} + \frac{4\alpha_{S^+}(1-\alpha_{S^+})(1-\beta^2)}{\pi^2 A^2} \end{aligned} \qquad (43)$$

## 3.2. Forced Vibrations

We collect terms of order $\varepsilon$ in equation (23):

$$D_0^2 X_1 + X_1 = -2\left((D_1 A)\cos(T_0 + \psi) - A(D_1\psi)\sin(T_0 + \psi)\right) + \sigma A \sin(T_0 + \psi)$$
$$- \gamma A \cos(T_0 + \psi) + \bar{A}\cos(T_0)$$
$$- (1-\beta^2)\left(C\sin(T_0+\psi) + \frac{8\alpha}{\pi^2}\left(\cos(\eta+\psi)\cos(T_0+\psi) + \sin(\eta+\psi)\sin(T_0+\psi)\right)\right) \quad (44)$$

The slow-flow equation describing the forced case is cast in the form:

$$D_1 A = -\frac{\gamma A}{2} - \frac{4\rho \alpha_{S^+}^2 (1-\beta^2)}{\pi^2 A} + \frac{\bar{A}}{2}\cos\psi$$
$$D_1\psi = D_1\psi = -\frac{1}{2}\left(\sigma - (1-\beta^2)\right) + \frac{4\alpha_{S^+}(1-\alpha_{S^+})(1-\beta^2)}{\pi^2 A^2} - \frac{\bar{A}}{2A}\sin\psi \quad (45)$$

Steady-state solutions corresponding to motion on the SIM are given by eliminating the derivatives in equation (45). Hence, the steady state solutions satisfy the following algebraic expression:

$$\pi^4 A_0^2 \bar{A}^2 = \left(\gamma\pi^2 A_0^2 - 8\rho\alpha_{S^+,0}^2(1-\beta^2)\right)^2 + \left(\pi^2\left(\sigma-(1-\beta^2)\right)A_0^2 - 8\alpha_{S^+,0}(1-\alpha_{S^+,0})(1-\beta^2)\right)^2$$
$$\alpha_{S^+,0} = \frac{1+\sqrt{(1+\rho^2)(A_0^2 - A_{min}^2)}}{1+\rho^2}; \quad A_{min} = \frac{\rho}{\sqrt{1+\rho^2}} \quad (46)$$

Here $A_0$ and $\alpha_{S^+,0}$ are steady-state amplitudes of $X$ and $f(T_0, T_1)$, respectively. $\rho_0$ is the steady state value of $\rho$, corresponding to $A_0$. Equation (46) is a fourth order polynomial with respect to the solution amplitude $A$. Under certain critical value of excitation amplitude $\bar{A}_b$ the polynomial does not have real roots, which means a periodic staeady-state solution is not possible.

# 4. Numerical results.

To assess the validity of approximate vibro-impact model with velocity-dependent restitution coefficient presented above, we compare the analytic predictions to numeric simulations based on the high-power potential and damping functions. The latter model is described by equations (9).

## 4.1. Free Vibrations

We compare between the approximate asymptotic solution amplitude $A(T_1)$ and numerical simulations of the full system equations (9) for $\bar{A} = 0$.

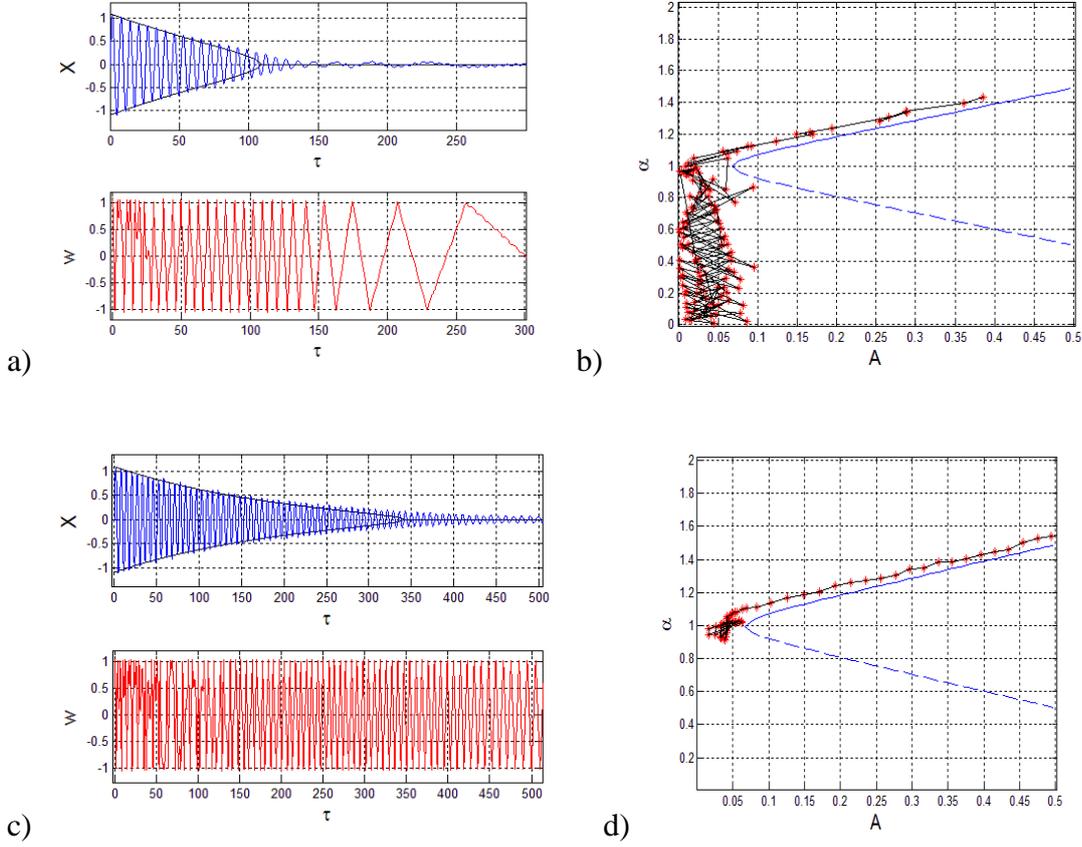

*Figure 9- a,c)Comparison between the responses time series of DOFs X (solid blue) and w (solid red) with and the analytical prediction of the slow-flow equationsX (solid black); b,d)Comparison between the system SIM (stable branch: solid-blue, unstable: dashed-blue) and the slow-flow motion(red dots connected with solid blach line) on the SIM stable branch, for initial conditions: $u_0 = \dot{v}_0 = 0, u_0 = 3, v_0 = -0.97$ and parameter values: $n = 50, \bar{p} = 0, \varepsilon = 0.05, \chi = 0.2, \lambda = 0.35$ a-b) $\beta = 0$; c-d) $\beta = 0.3$*

As one can see in Figure 9, there is a good agreement between numerical results and analytical prediction, both in the absence of linear component ($\beta = 0$) and for $\beta = 0.3$. The significant difference is that, unlikely the free VI particle, for nonzero β the IP keeps oscillating with its natural frequency after escaping from the SIM stable branch. For $\beta > 0$, after escaping the impact regime, there is no damping mechanism for the DOF *w*, and as a result, the system keeps oscillating withconstant amplitude. Moreover, it is noteworthy that as parameter $\beta$ increases, the asymptotic approximation losses its validity.

One can see in Figure 9(a)-(d) that as long as the system is captured into 1:1 resonance state ((a)-(b) until $\tau \approx 100$, and (c)-(d) until $\tau \approx 350$) the analytical prediction is in a good agreement with the numerical simulation envelope. Superposition of the SIM and the slow-flow numerical simulation is demonstrated in Figure 9(c). The system moves on the SIM stable branch only in the state od1:1 resonance. After approaching the fold, the dynamical flow leaves the SIM. . This is the explanation of the red dots in Figure 9(d) that are far from the branch. The amplitude remains in the vicinity of $A = A_{\min}$, since no additional damping mechanism exists in the model, besides the impacts.

## 4.2 Forced Vibrations

### 4.2.1. Response to a single frequency periodic excitation

There are three possible response regimes for the constant-amplitude and single-frequency external forcing. The first regime is a resonant periodic oscillations with constant amplitude (Figure 10). The particle performs resonant impacts in approximately constant pace. Consequently, the amount of energy in the primary system is almost constant. .

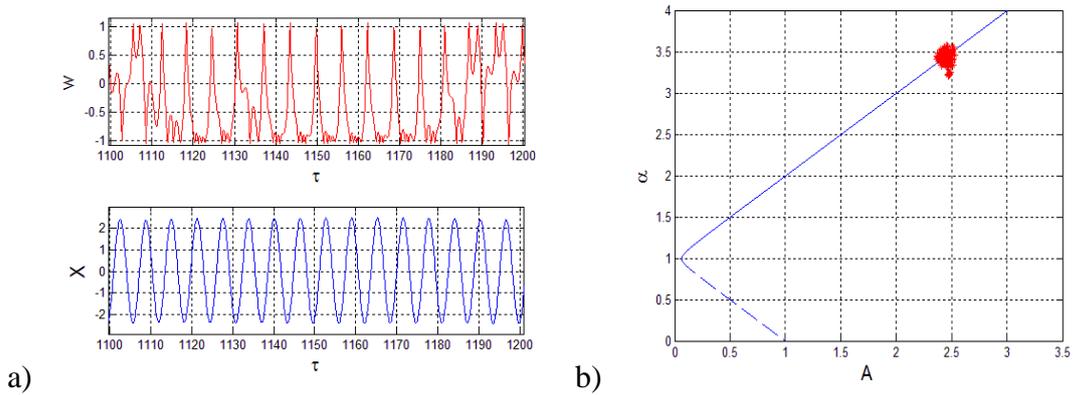

a) b)

*Figure 10-Forced-system steady-state resonant periodic regime*
$\varepsilon = 0.05, \gamma = 0.2, \lambda = 0.35, \chi = 0.2, \bar{A} = 3, \sigma = 1, n = 50, \bar{p} = 0, \beta = 0$, *Initial conditions:*
$u_0 = 0, \dot{u}_0 = 3, v_0 = -0.97, \dot{v}_0 = 0$, *a) time history responses in time interval* $\tau \in [1100, 1200]$, *b) system motion on SIM.*

The second regime is non-resonant steady-state periodic oscillation. They take place when the system isnot captured into the 1:1 resonance and the does not reach the SIM stable branch (Figure 11). In this case, when the particle is coupled with the linear spring ($\beta > 0$, Figure 11(b)) it oscillates in its natural frequency, impacts do not take place and as a result, the main structure vibrates with almost a constant amplitude, which do not belong to the SIM stable branch.

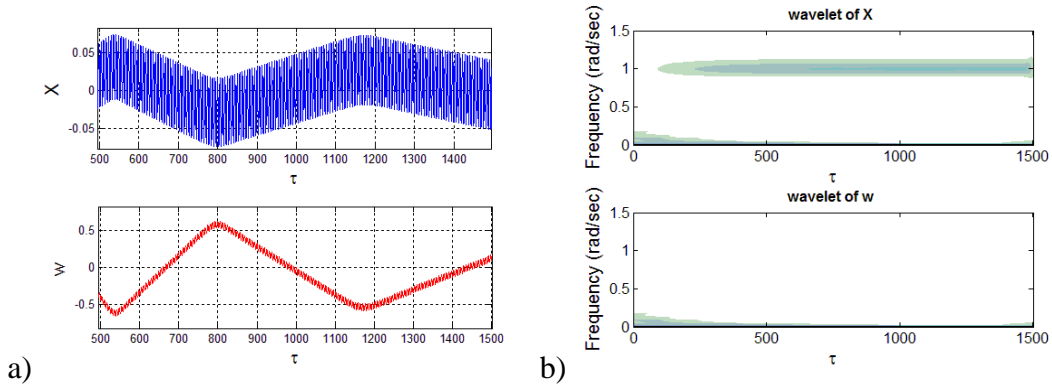

a) b)

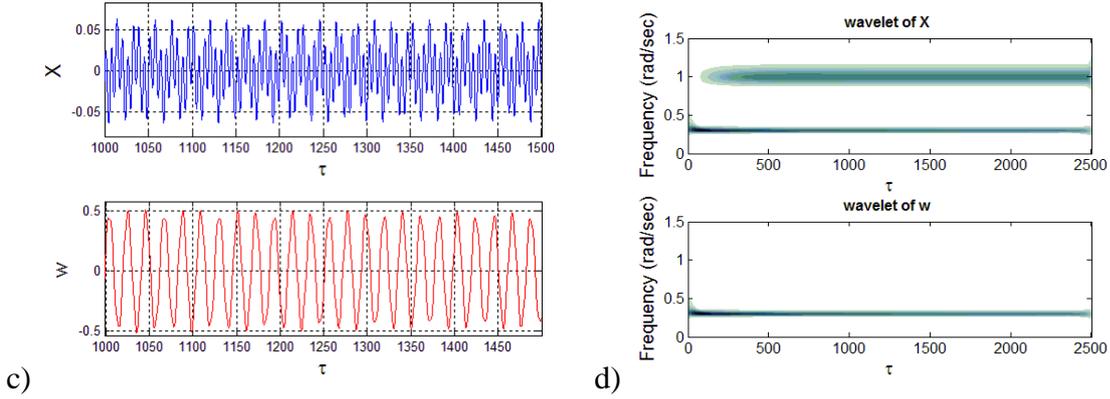

c)     d)

*Figure 11- Forced-system steady-state non-resonant periodic regime*

$\varepsilon = 0.05, \gamma = 0.2, \lambda = 0.35, \chi = 0.2, \overline{A} = 0.01, \sigma = 1, n = 6, \overline{p} = 0$, *Initial conditions:*
$u_0 = 0, \dot{u}_0 = 0, v_0 = -0.97, \dot{v}_0 = 0$; *time histories and corresponding wavelet plots for: a,b)* $\beta = 0$, *c,d)*
$\beta = 0.3$.

The third regime is strongly modulated oscillations, which correspond to intermittent resonance capture events (Figure 12(a,b)). Between these events, the system approaches the SIM and $X$ amplitudee grows. Then the transient resonance capture takes place, the system reaches the SIM stable branch, the amplitude decays rapidly until the solution amplitude reaches bifurcation value of $A_{min}$, and so on.

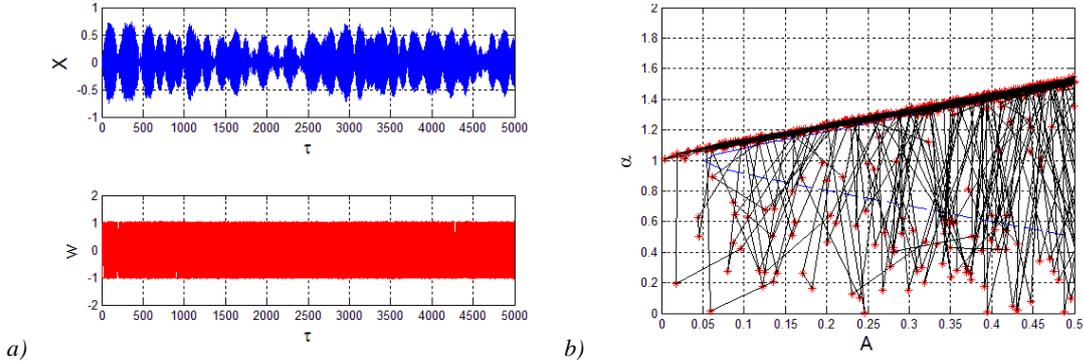

a)     b)

*Figure 12- CSMR regime of the forced system*

$\varepsilon = 0.05, \gamma = 0.2, \lambda = 0.35, \chi = 0.2, \overline{A} = 0.5, \sigma = 1, n = 6, \overline{p} = 0, \beta = 0.3$, *Initial conditions:*
$u_0 = 0, \dot{u}_0 = 0, v_0 = -0.97, \dot{v}_0 = 0$ *a) time history responses in time interval* $\tau \in [0, 5000]$, *b) system motion on SIM.*

For certain parameter sets, transition between different regimes is obtained. For instance, one can see in Figure 13 transition from theCSMR regime to the steady-state periodic oscillations. This transition is expressed by motionalong the SIM stable branch until attraction to the stationary point.

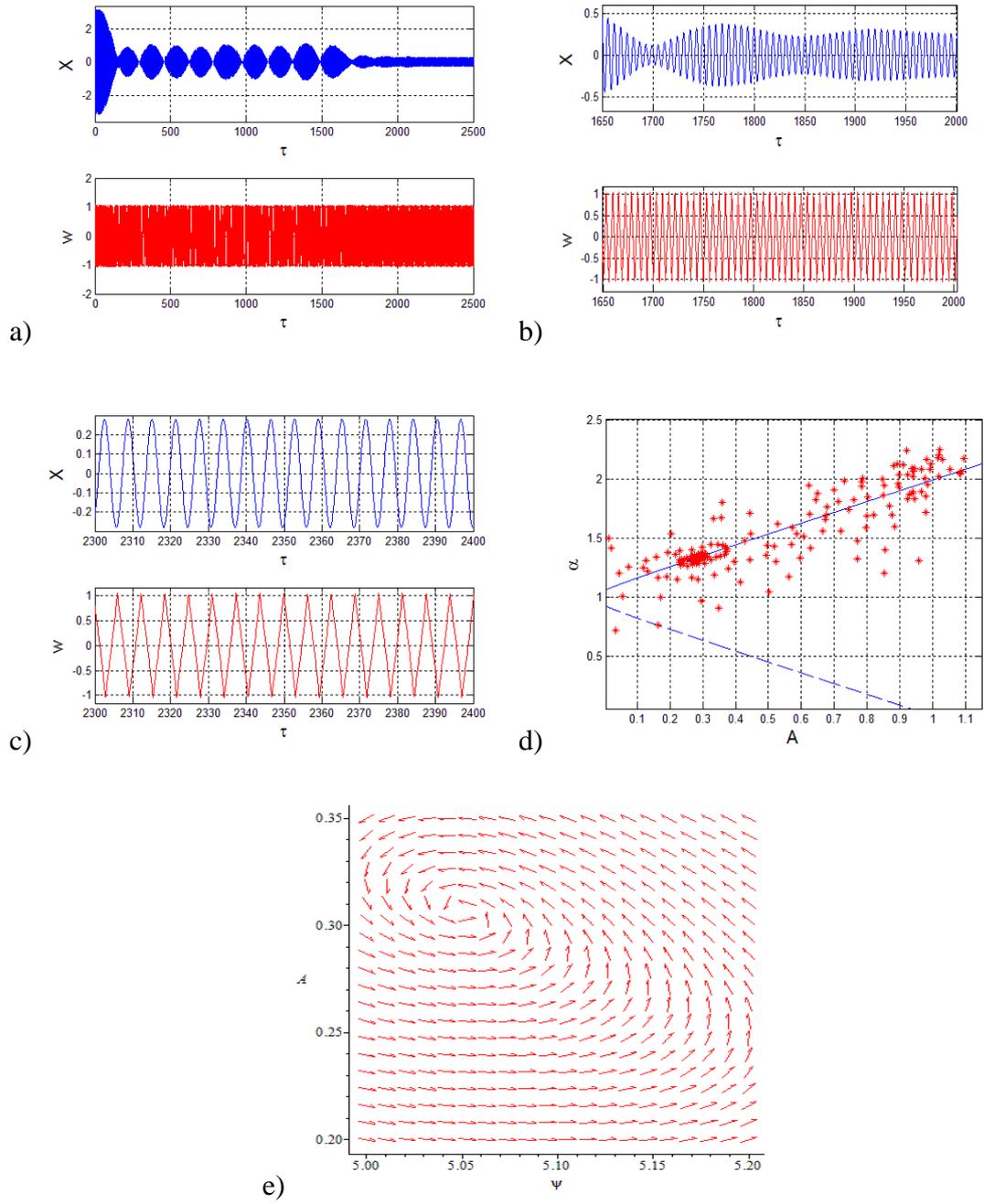

*Figure 13-* $\varepsilon = 0.05, \gamma = 0.2, \lambda = 0.35, \chi = 0.2, \bar{A} = 1, \sigma = 1, n = 50, \bar{p} = 0, \beta = 0.3$, *Initial conditions:* $u_0 = 0, \dot{u}_0 = 3, v_0 = -0.97, \dot{v}_0 = 0$ *time histories and dynamical flow plots,*

### 4.2.2. Narrow-band random excitation

In previous sections we analyzed the system response to the single-frequency periodic excitation. In the following section we analyze the modification of the system response, if small-amplitude zero-mean white noise is added to the periodic excitation. i. The random excitation is defined as follows:

$$F(\tau) = \varepsilon A \cos(\tau) + \varepsilon F_w\left(\mu = 0, \bar{\sigma}^2\right) \tag{47}$$

Here $F_w\left(\mu = 0, \bar{\sigma}^2\right)$ is the additive zero-mean white Gaussian noise with standard deviation of $\bar{\sigma}$. The white noise signal is generated by computing a random number and multiplying it by the standard deviation $\bar{\sigma}$. This excitation force is exemplified in Figure 14, for periodic part with $A = 1$ and standard deviation $\bar{\sigma} = 4$.

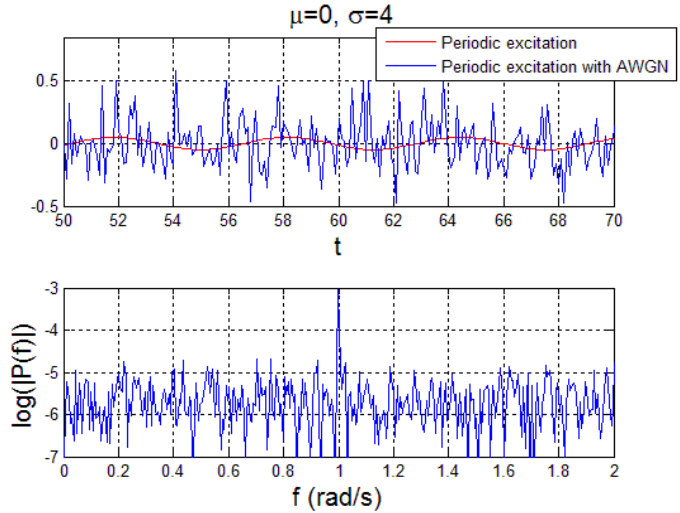

*Figure 14- Periodic and zero-mean additional white Gaussian Random forcing excitation for parameter set: $\bar{\sigma} = 4, \varepsilon = 0.05, \bar{A} = 1$, a)excitation time history; b)Fast Fourier Transform(FFT).*

Response in Figure 15 is computed or the same parameters and as for Figure 10, except additional Gaussian white noise..

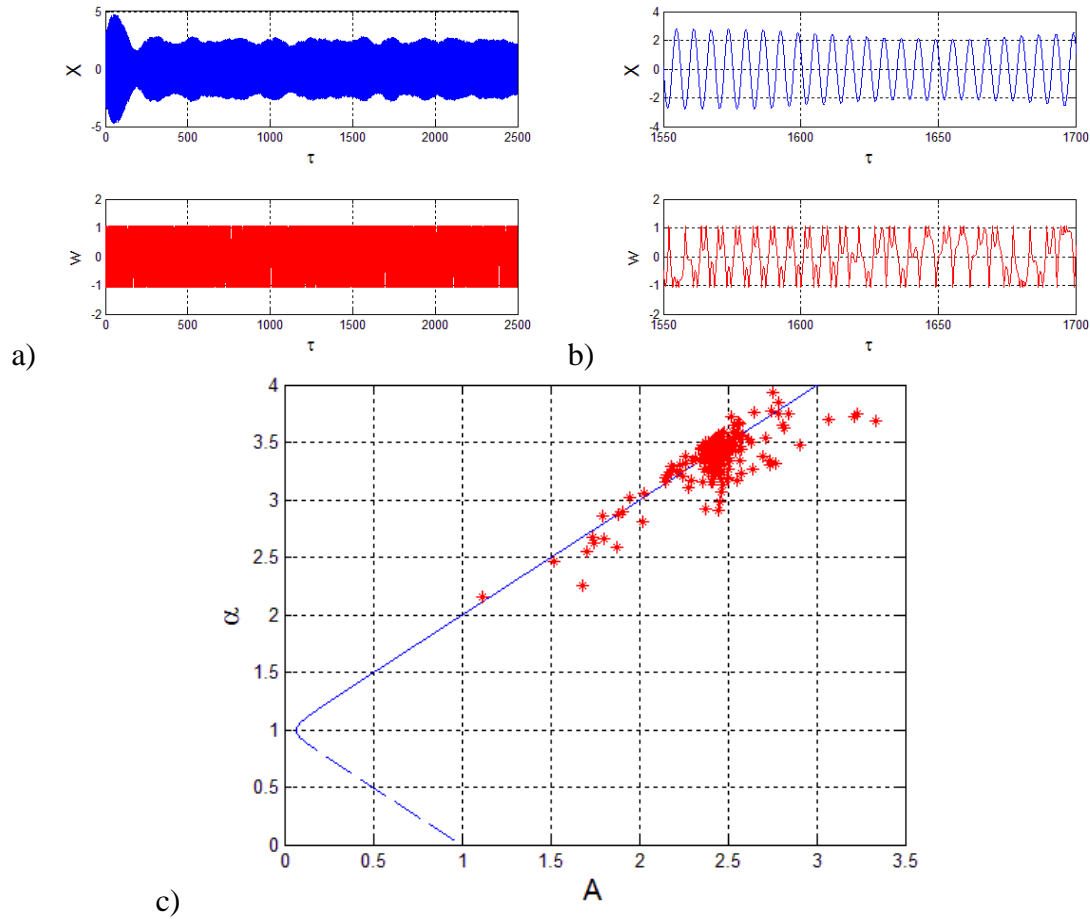

a)
b)
c)

*Figure 15-Times histories and system motion on the SIM for parameters:*
$\varepsilon = 0.05, \gamma = 0.2, \lambda = 0.35, \chi = 0.2, \bar{A} = 3, \sigma = 1, n = 50, \bar{p} = 0, \beta = 0$, *Initial conditions:*
$u_0 = 0, \dot{u}_0 = 3, v_0 = -0.97, \dot{v}_0 = 0$.

As one can see, there is no major qualitative difference due to the added noise. However, it may lead to perturbations of the motion on the SIM. .. Moreover, if the response amplitude is close to the bifurcation point, the noise will facilitate formation of the regime similar to the CSMR.

## 5. Concluding remarks

Exploration of dynamical responses of the system with nonlinear liquid sloshing is traditionally based on reduced-order model that involves high-power force and damping functions. One can see that these responses can be successfully described with the help of even simpler effective vibro-impact model with velocity-dependent restitution coefficient. The simplification is so significant, that the latter model yields to analytic exploration by multiple-scale approach.

The paper checks the relationship between two levels of the reduced-order modelling. However, of course, it does not check the relevance of any of these models for description of the actual sloshing dynamics. In the same time, it seems that the vibro-impact model is more suitable for the comparison with experiments or extended numerical simulations. It qualitatively predicts a small amount of possible response regimes; these predictions are verifiable. Besides, the quantitative side of the treatment allows determination and fitting of the phenomenological parameters.

The authors are very grateful to Israel Science Foundation (grant 838/13) for financial support.